# A Comparison of Six Methods for Stabilizing Population Dynamics

**Sudipta Tung[1], Abhishek Mishra[2] and Sutirth Dey[3]***

**Affiliations:**

Population Biology Laboratory, Biology Division, Indian Institute of Science Education and Research-Pune, Dr Homi Bhabha Road, Pune, Maharashtra, India, 411 008

*Correspondence to Email: s.dey@iiserpune.ac.in

[1] **sudipta.tung@students.iiserpune.ac.in**

[2] **amishra@students.iiserpune.ac.in**

[3] **s.dey@iiserpune.ac.in**

**Name and address of the corresponding author:**

Sutirth Dey
Assistant Professor, Biology Division
Indian Institute of Science Education and Research
Dr Homi Bhabha Road,
Pune, Maharashtra, India 411 008
Tel: +91-20-25908054

## Abstract

Over the last two decades, several methods have been proposed for stabilizing the dynamics of biological populations. However, these methods have typically been evaluated using different population dynamics models and in the context of very different concepts of stability, which makes it difficult to compare their relative efficiencies. Moreover, since the dynamics of populations are dependent on the life-history of the species and its environment, it is conceivable that the stabilizing effects of control methods would also be affected by such factors, a complication that has typically not been investigated. In this study we compare six different control methods with respect to their efficiency at inducing a common level of enhancement (defined as 50% increase) for two kinds of stability (constancy and persistence) under four different life history/ environment combinations. Since these methods have been analytically studied elsewhere, we concentrate on an intuitive understanding of realistic simulations incorporating noise, extinction probability and lattice effect. We show that for these six methods, even when the magnitude of stabilization attained is the same, other aspects of the dynamics like population size distribution can be very different. Consequently, correlated aspects of stability, like the amount of persistence for a given degree of constancy stability (and vice versa) or the corresponding effective population size (a measure of resistance to genetic drift) vary widely among the methods. Moreover, the number of organisms needed to be added or removed to attain similar levels of stabilization also varies for these methods, a fact that has economic implications. Finally, we compare the relative efficiency of these methods through a composite index of various stability related measures. Our results suggest that restocking to a constant lower threshold seems to be the optimal method under most conditions, with the recently proposed Adaptive Limiter Control (ALC) being a close second.

# Introduction

Since the seminal work of Ott, Grebogy and Yorke [1], a large number of methods have been proposed to stabilize the dynamics of unstable non-linear systems (for reviews see [2-4]). Many of these methods work by manipulating the parameters of the system in real time, such that the trajectory of the system can be stabilized to the desired kind of dynamics (stable point or limit cycles of appropriate periodicity). However, such methods are unsuitable for controlling real biological populations in which the precise equations governing the system are typically unknown and parameters (e.g. intrinsic growth rate, carrying capacity etc.) can only be estimated *a posteriori*, through model-fitting. Control of biological populations is more easily achieved through methods that stabilize the dynamics through perturbations to the state variable, (i.e. the population size) and require relatively less system-specific information. Over the last two decades, many such methods have been proposed [5-9] and at least a few of them have also been empirically verified [9-12].

This proliferation of biologically relevant control methods has created some interesting problems of its own. In ecology, there are multiple notions about the concept of stability [13] and ideally one would not like to opt for a method that enhances one kind of stability (say reduction in fluctuation in population size) at the cost of another (say long term persistence). However, studies on control methods often focus on enhancement of only one type of stability, without investigating how other aspects of the dynamics get affected (e.g. [5,8,14]). Recent empirical studies indicate that induction of one kind of stability may [9] or may not [15] translate into the enhancement of other kinds of stability. Therefore it is important to investigate how different control methods affect multiple kinds of stability simultaneously.

Such comparisons can be quite complex as most theoretical studies employ different models of population growth and evaluate the efficacies of the control methods in different parameter ranges, some of which can even be biologically unrealistic. Thus, for meaningful comparison, these methods need to be investigated under common conditions, i.e. for the same model and similar levels of enhancement of stability. Moreover, since it has been shown that the effects of perturbation can vary depending on the intrinsic growth rates or the environment of the population [16], it is conceivable that the efficacy of control methods can also be affected by these factors. Thus, any comparison of the control methods also needs to take into account multiple combinations of intrinsic growth rate and carrying capacity values. Finally, any real world scenario typically involves an economic component [7], which might play a significant role in deciding which control method is best suited to a given scenario. Our study aims to compare the performance of six well-known control methods in population dynamics under the above-mentioned set of conditions.

Here, owing to logistic constraints, we restrict our analyses to six control methods which were selected based on two criteria. Our primary selection criterion was the relative ease with which the methods could be implemented in real, biological populations. This ruled out some of the well-known, empirically verified control methods that require extensive knowledge of the equations governing the system and the corresponding parameter values [11,17]. Our second criterion was the extent of information already available about the control methods in the population dynamics literature. Barring one (Both Limiter Control, see below), for which we found no prior reference in the literature, all the methods that we chose have been extensively investigated both analytically and numerically, and have been shown to be robust to at least some degree of noise. We realize that there might be other control methods that fit these two criteria and therefore do not claim that our coverage is comprehensive.

The mathematical expressions for the six control methods and the corresponding ranges investigated in the exploratory analysis are given in Table 1. Here we present a brief description of how these methods stabilize population dynamics. Among the six, constant pinning (CP) is perhaps the most well studied [8,18-19] and involves the influx of a constant number of individuals (from some external source) into the population in every generation. In its general form, CP involves both immigration and emigration from a population [18], but here we concentrate solely on immigration which has been shown to enhance stability for populations governed by the Ricker [20] dynamics [21]. The reason for this is best understood graphically. For models that have single-humped first-return maps with at most one inflection point to the right of the maximum, the nature of the dynamics depends upon how negative the slope of the first-return map is at the point where it intersects the 45° line. Since constant immigration shifts the entire return map upwards (see Fig 2 of [22]), the slope at this point is reduced, which can convert chaotic dynamics into limit cycles or even stable points, depending upon the magnitude of the reduction [18]. For those models, such as the logistic, where moving up the first-return map increases the slope at the intersection point with the 45° line, CP destabilizes the dynamics by making it more complex [18]. Biologically, CP creates a "floor" and does not allow the population to hit values below the constant immigration threshold. This method has been empirically demonstrated to reduce fluctuations in sizes for spatially-unstructured [16] but not spatially structured populations [12].

One of the issues with constant pinning is that the population sizes are augmented even when they are not low. This problem is avoided with the so called hard 'limiter control from below' [7], or Lower Limiter Control (LLC) in this study, which prescribes that each time the population size falls below a pre-determined lower threshold, it is brought back to that value through restocking. Graphically, LLC truncates some part of the lower end of the return map, which in turn makes part of the upper end unavailable to the system [7]. This constrains the range of values that the system can take, finally leading to stabilization of the dynamics [23]. A similar truncation of the return map can also be obtained by bringing the population size back to a given upper threshold (through culling) every time it crosses the threshold [24]. This is the control strategy known as hard 'limiter control from above' [7] and referred to as Upper Limiter Control (ULC) in this study. A logical extension of the LLC and the ULC scheme is to combine both and bring the population to a lower or a upper threshold each time the size goes respectively below or above those limits. This is what we term as Both Limiter Control (BLC), a scheme that has not been proposed earlier to the best of our knowledge.

One of the drawbacks of all the four methods described above is that the value of the control parameter (i.e. the various thresholds, or the fixed number of immigrants for CP) has to be set *a priori*, which requires some knowledge of the system in terms of both growth rates and carrying capacity, which are assumed to be unchanging. More realistic control methods avoid this assumption by setting the magnitude of the control parameter proportional to the current population size [14,25]. We therefore decided to also investigate one such recently proposed method called Adaptive Limiter Control or ALC which has been theoretically and empirically demonstrated to reduce the fluctuation in size of both spatially structured and un-structured populations [9]. Unlike the previous methods, ALC is not capable of turning a chaotic trajectory into a stable point equilibrium, but "traps" the dynamics into a region around the carrying capacity [26]. The range of this trapping region is determined by both the growth rate and the value of the controlling parameter [26] and the nature of the dynamics can be either limit cycles or chaos [9]. For situations where it is desirable to guide the system towards a particular value of the state variable rather than a range of values, the so-called Target-Oriented Control or TOC [6] is of greater use. In this method, the system is guided towards a particular target value (set *a priori* based on whatever is the desired usage) by introducing individuals whenever the population size is below the target and removing individuals when the population size is above the target. The number of individuals to be added or removed is proportional to the difference between the target and the current value of the population size. It has been analytically shown that for high enough values of this proportion, TOC will always lead the system to a positive stable point [27].

In this study, we compare the efficiency of these six methods in inducing a common level (50%) of reduction in either fluctuation in population sizes or extinction frequencies under four different life history/ environment combinations. Since these control methods have been analytically investigated elsewhere (see above for details), we focus on an intuitive understanding of how these methods change the distributions of population sizes over time, thereby affecting fluctuations, extinctions, effective population sizes and the amount of perturbation required to attain a defined stability goal. We use Ricker model–based, biologically realistic simulations incorporating parameter noise, stochastic extinctions and lattice effect (*sensu* [28]). We show that for these six methods, even when the degree of stability attained is similar, the resulting population size distributions can be very different. Consequently, for a given degree of stability attained, the correlated features of dynamics (e.g. extinction probability, ability to resist genetic drift etc.) vary widely among the methods. The magnitude of perturbation needed to attain similar levels of stability was also different for these methods, which is likely to have economic consequences. Finally, we computed a composite index of various stability related measures to compare the relative efficiency of these methods. Our results suggest that under most conditions, Lower Limiter Control, i.e. restocking to a constant lower threshold, seems to be the optimal strategy.

## Methods

*Population growth model*

We used the well-studied Ricker model [20] for representing the population dynamics. This map is given as $N_{t+1}=N_t*exp(r*(1-N_t/K))$, where *r, K* and $N_t$ denote intrinsic growth rate, carrying capacity and population size at time *t*, respectively. Due to its simplicity and lack of species specific parameters, this model has been extensively investigated theoretically [29]. First principle derivation suggests that populations exhibiting scramble competition and random spatial distribution should exhibit Ricker dynamics [30]. Since populations of several species are expected to exhibit these properties, the model has been widely used to describe the dynamics of, *inter alia,* bacteria [31], fungi [32], ciliates [33], insects [34-35] and fishes [20,36]. Therefore, simulation results obtained here from this model are expected to be broadly applicable across a large number of taxa.

*Life history / environmental regimes*

The aim of this study was to compare the efficacy of various ecologically meaningful control algorithms under biologically relevant common conditions. However, the intrinsic growth rates and carrying capacities of species under different environments vary widely. Thus, it is conceivable that control algorithms might differ in terms of their efficacy and/or cost of implementation under various environments. Since it is not possible to represent every growth rate-carrying capacity combination, we arbitrarily chose two levels of intrinsic growth rate *r* (Low = 2.8 and High = 4.0) crossed with two levels of carrying capacity *K* (Low = 60 and High = 300). Thus, we investigate four combinations HrHk, HrLk, LrHk and LrLk, where HrHk denotes a combination of *r* = 4.0 and *K* = 300 and so on. Since we aimed to study the stabilizing effects of the control methods, we explicitly chose *r*-values that led to extinctions and large amplitude oscillations in population sizes in the unperturbed cases. The value of $r = 2.8$ is representative of the growth rates of some laboratory insect populations [34-35] whereas the value of $r = 4.0$ is well within the limits of Ricker growth rates estimated from natural populations of fishes [36]. In the context of the Ricker model, $r = 2.8$ represents a value just after the onset of chaotic dynamics (which happens at $r = 2.697$) while $r = 4.0$ represents highly chaotic dynamics [29]. The value of Lk (= 60) was chosen to be similar to a previous analysis [9] while Hk was arbitrarily chosen as 5 times Lk. Thus, the four regimes represented dynamics that are well within the ranges of real biological populations in the laboratory and nature. In all cases, the initial population size ($N_0$) was taken to be 20.

*Transients and lattice effect*

Most studies in the control literature tend to investigate the dynamics of the system under steady states. However, many factors that determine the dynamics of a population (e.g. the environment of its habitat or the distributions of various life-history traits) are unlikely to remain constant for long [37]. Moreover, due to the time scales involved, most available ecological time series are short and, therefore, unsuitable for checking the predictions on steady-state dynamics. Therefore, following earlier work [9,26] we explicitly concentrated on the transient dynamics by restricting our simulations to the first 50 iterations. Moreover, we rounded off the number of organisms and the magnitude of the perturbations to the nearest integer values. This accounted for the fact that real organisms always come in integer numbers (lattice effect, [28]), which is known to significantly affect the dynamics in simulation studies [38].

*Extinctions and resets*

In the absence of lattice effect, a Ricker-generated time-series can never take a zero value (i.e. become extinct) when initiated with non-zero population sizes. However, in simulations incorporating the lattice-effect, the Ricker dynamics permits extinctions whenever the population size goes below 0.5. We call this kind of extinction as Lattice Effect Extinction (LEE) which happens when $INT[f(N_{t-1})] = 0$, where INT is a function rounding off the population size to the nearest integer. Following previous empirical studies [9,16,34], we also incorporated extinction due to demographic stochasticity (EDS) in the form of a 50% probability of extinction whenever the population size went below four. Mathematically, this can be represented as $P(N_t = 0 \mid N_t' <= 4)) = 0.5$ where $N_t$ denotes the population size in generation *t* after the extinction step and $N_t'$ denotes the population size in generation *t* after the application of the Ricker model on $N_{t-1}$ (i.e. $f(N_{t-1})$). Biologically, EDS occurs due to chance realizations of probabilistic events in a population, e.g. when all the members of a population are of the same sex, or are infertile, or fail to reach adulthood [39]. Prior simulations using this level of EDS have been seen to give good fits to trends from experimental time series of laboratory populations of *Drosophila melanogaster* [9,34]. For unperturbed populations and those controlled by ULC, the population size was reset to a value of eight after an extinction event [9,34]. In the case of all other control methods, the extinct populations were automatically reset by the respective control schemes.

For representational purposes, we scaled the extinction probability for all control methods in a give regime by the average extinction probability of the unperturbed population (Fig 4A, Fig S10A) in that regime. This allowed us to directly compare the efficacies of the methods across regimes.

*Stochasticity and replication*

Since noise is known to have a major impact on the dynamics of perturbed populations [12], we incorporated noise in both *r* and *K* in each iteration of the simulations. Thus, the stochastic intrinsic growth rate was given as *r* + *ε* (where $\varepsilon \in U[-0.2, 0.2]$) and the stochastic carrying capacity as *λK* (where $\lambda \in U[0.9, 1.1]$). All simulations were repeated 50 times and the corresponding mean values and standard errors around the mean are reported here. The small error bars indicated that 50 replicates were enough for the purpose of our study. However, we repeated a subset of our simulations with 1000 replicates and found no qualitative difference with our results (data not shown).

*Effort magnitude and Effective Population Size ($N_e$)*

The cost of implementation of the six control methods was quantified as the effort magnitude defined as the average number of individuals added or removed per generation from the population in order to attain the desired control level [7]. It can be computed as:

$$\frac{1}{T \times \overline{\overline{N}}} \times \sum_{t=1}^{T} |b_t - a_t|,$$

where $b_t$ and $a_t$ are the population sizes before and after perturbation in the $t^{th}$ generation and $\overline{N}$ and *T* denote the average population size and length of the time series respectively (Sah et al. 2013). Note that effort magnitude defined this way is dimensionless, being a fraction of the corresponding mean population size, thus permitting direct comparisons across different regimes. Following earlier work [7,9], we assumed that lower values of effort mean less expense and therefore are economically more viable.

Effective Population Size ($N_e$) was quantified as the harmonic mean of the post-perturbation time series [40]:

$$N_e = \frac{T}{\sum_{t=1}^{T} \frac{1}{N_t}}$$

where, $N_t$ is the population size in generation $t$ and $T$ is the length of the time series. For representational purposes, we scaled the $N_e$ for all control methods in a give regime by the average $N_e$ of the unperturbed population (Fig 4C, Fig S10C) in that regime. This allowed us to directly compare the efficacies of the methods across regimes.

*Measures of stability*

Populations with relatively large amplitude of fluctuations in population size are considered to have lower 'constancy' stability and vice versa [13]. The constancy stability of the simulated time series was quantified using the Fluctuation index, FI [34] which represents the average one-step fluctuation in population size, scaled by the population mean:

$$FI = \frac{1}{T \times \overline{N}} \times \sum_{t=0}^{T-1} |N_{t+1} - N_t|$$

where, $\overline{N}$ is the average population size, $N_t$ is the population size in generation $t$ and $T$ is the length of the time series. Persistence stability is the converse of the extinction probability of a population [13]. Mathematically, *extinction probability* $= \frac{E}{T}$, where $E$ represents the total number of extinction events during T iterations.

*Composite Performance Score*

One of the aims of this study was to compare the overall performance of the six control methods against each other. Therefore we devised a composite performance score which combined the relative performance of each control method w.r.t FI, extinction probability (EP), effort magnitude (EM) and $N_e$. Since the ranges of values taken by these quantities are very different, we normalized each quantity in each regime, by the highest value in that regime. So for example, in the HrLk regime, since ULC had the highest effort magnitude (Fig 4B), the corresponding values for all the methods were divided by the effort magnitude of ULC. This normalizes all values to a scale of 0 to 1, with 0 being the best performance and 1 being the worst. This works well for fluctuation index. extinction probability and effort magnitude where lower values are more desirable. Since higher values of $N_e$ are more preferable, we subtracted the normalized $N_e$ values from 1. The composite performance score for each method in a given regime was simply the sum of the three normalized scores, which implies equal weightage to each score. In other words, for 50% reduction in FI, the performance score is given as $EM' + EP' + (1- N_e')$ and for 50% reduction in extinction probability, the corresponding expression is $EM' + FI' + (1- N_e')$, where prime denotes the normalization operation. Clearly lower values of this score for a given method in a particular scenario indicate better performance than higher values.

## Results and Discussion

**Dynamics of unperturbed populations**

We first explored the dynamics of the four regimes in the absence of any perturbation. The two Hr regimes (i.e. HrHk and HrLk) showed higher FI (Fig 1A) and greater extinction probability (Fig 1B) compared to the corresponding Lr regimes. This difference between Hr and Lr is not surprising in terms of constancy, as a higher value of *r* implies greater amplitude of fluctuation in the time series and hence a greater FI. A larger FI is also expected to increase the frequency with which a population crashes to very low values, and thus reduce persistence. Interestingly, although HrLk had the same *r* and a marginally higher FI than HrHk (Fig 1A), the corresponding extinction probability was approximately double (Fig 1B). This is intuitive since increasing the habitat capacity of a population (which is analogous to *k* here) is expected to increase the time to extinction [41]. This also highlights the point that constancy and persistence of a population need not necessarily have a simple relationship and it is risky to try and extrapolate one from the other, an observation that seems to have a fair bit of empirical support of late [9,15-16].

Another important factor that increases the chance of population extinction is the erosion of genetic diversity through genetic drift. Although the importance of drift is well recognized in evolutionary [42] and conservation biology literature [43], most studies in population control have ignored the effects of a control method on genetic drift. Here we investigate this phenomenon by studying the effective population size $N_e$ (Fig 1C), which is defined as the corresponding size of an ideal population that has an equivalent rate of loss of heterozygosity as the population under study [40]. The rate of loss of heterozygosity increases with decrease in population sizes or bottlenecks [44]. More interestingly, even the rate of generation of genetic variation, as measured by the average mutation rate, scales negatively with $N_e$ among the major phylogenetic groupings [45]. This suggests that higher values of $N_e$ are more desirable since they are expected to cause lesser loss of genetic variation. When the $N_e$ of a series is measured as the corresponding harmonic mean, it tends to get much more negatively affected by the presence of low values than the corresponding arithmetic mean. This suggests that the two Hr regimes (HrHk and HrLk) are expected to have much lower values of $N_e$ compared to the two Lr regimes. However, the $N_e$ of HrHk and HrLk were found to be roughly similar to that of LrLk. This discrepancy is explained by the fact that both Hr regimes were marked by large number of extinctions (Fig 1B) which were followed by resets to a value of eight (See Section *Methods: Extinctions and resets*). Consequently, all the zero values in the population sizes in these two regimes were replaced by eight, which significantly increased the $N_e$. We verified this line of reasoning by simulating the unmodified Ricker model (i.e. in the absence of lattice effect, noise or extinction) and found that the $N_e$ under HrHk and HrLk regimes were actually orders of magnitude less compared to the two Lr regimes. Therefore, the high values of $N_e$ in the two Hr regimes should be considered an artefact of the extinction-reset protocol.

Thus, even though the dynamics of all four regimes are chaotic, we find important differences in their properties in the absence of any perturbation. We next investigated how the six different control methods affected these properties.

*Exploratory Analysis*

Literature survey indicated that given high enough values of the corresponding control parameter, most of the methods are expected to turn chaotic dynamics into stable points or low amplitude limit cycles. However, the relative efficiencies of these methods are hard to compare since they have typically been studied in the context of different kinds of stability properties. For example, TOC has only been looked at in the context of ameliorating chaotic dynamics to a stable point [6], where as ALC does not ameliorate chaos to begin with [9,26]. Therefore in order to make the comparisons meaningful, we began by exploring the efficacy of all six methods in enhancing the constancy and persistence stability over a wide but biologically / realistically meaningful parameter range (Fig S1-S6 in SOM). Here, we stress upon the meaningfulness of the parameter ranges, since some times the methods perform the best under parameter values that are realistically unfeasible or biologically undesirable. For example, for CP, we only considered number of immigrants up to $k$, since we feel that perturbation sizes greater than the carrying capacity are impractical. Similarly, for cases where the population size is not allowed to go below a fixed threshold (Lower Limiter Control or LLC), we only considered values less than the carrying capacity, since setting a lower threshold above $k$ is biologically unrealistic.

From Fig S1-S6, it is clear that although there were differences in terms of their performance, all six methods were able to reduce the fluctuation index under all four regimes. To create a common platform for comparison, we extracted the values of the control parameters from Fig S1-S6 that lead to a 50% reduction in Fluctuation Index (FI) and Extinction Probability w.r.t the unperturbed dynamics as in Fig 1. Table S1 summarizes these values of control parameters. We then compared these six methods in terms of their effort magnitude, resulting effective population size ($N_e$) and the corresponding extinction probability or fluctuation index. Finally, we combined the performance of all these methods to come up with a common score that would help us to choose one method over another in a given scenario. It should be noted here that we do not seek to establish mathematically rigorous results on how these methods actually stabilize the dynamics. We have explicitly chosen methods for which such information already exists (see Introduction for the relevant references). The current study aims to create an intuitive understanding of how the different control methods alter the distributions of pre- and post-perturbation population sizes. The reasons for focussing on population size distributions are two-fold. Firstly, although it has been shown that in response to a particular control method, the pre- and post-perturbation population sizes can have very different dynamics [27], the phenomenon has not been explored for other methods. Secondly, as we demonstrate here, various aspects of population dynamics like effort magnitude and the probability of extinction can be ultimately thought of as an interaction between the pre-and post- perturbation population sizes.

*50% reduction in FI: Persistence stability*

Although extinction has been extensively studied theoretically and empirically ([46] and references therein), few studies on control methods have explicitly considered enhancement of persistence (although see [9,16]). This is perhaps because many studies on controlling single species biological populations come from the tradition of chaos control in non-linear dynamics, and primarily focus on the attainment of simpler dynamics [6,8,18] or reducing the variation in population size, i.e. constancy [7]. Therefore, we first investigated how achieving a given level of constancy affects the corresponding persistence stability.

For a similar level of reduction in FI (i.e. 50% of the unperturbed population), the corresponding population size distributions were found to be very different for the six methods, particularly in the two Hr regimes (Fig 2 and 3). The lower percentiles (5$^{th}$, 10$^{th}$ and 25$^{th}$) of post-perturbation population sizes are higher than the unperturbed for all the control methods in all four regimes (Fig 2). This is not surprising for all methods that include a restocking component (i.e. except ULC). In the case of ULC keeps the lower percentiles are high due to prevention of crashes from high numbers in the previous generation. This becomes apparent from the distribution of the pre-perturbation population sizes (Fig 3) which shows that ULC maintains the lower percentiles of population sizes higher than all the other methods. This leads to the prediction that ULC is expected to have the least probability of extinction for this level of enhancement of constancy and indeed we found no extinctions at all under ULC in any of the four regimes (Fig 4A). This result apparently contradicts an earlier study that found that the imposition of an upper limiter is actually expected to increase the extinction probability of a population [24]. This discrepancy is due to the fact that the earlier study assumed density-independent, randomly distributed growth rates, due to which the chances of an extinction in the next generation (*t+1*) increased monotonically with any kind of reduction in population size in generation *t* [24]. However, in this study, we explicitly assumed growth rates to be density-dependent, as a result of which, the extinction probability in *t +1* actually goes down with reduction in population size in *t*. As long as the ULC threshold is not set at extremely low levels (i.e. so low as to become extinction pre-images) and there are no Allee effects (which was not considered in this study), reducing the

ULC upper limit is expected to promote persistence. This discrepancy between the results of the two studies highlights that the effects of a particular control method can be conditional upon the nature (i.e. density-dependent vs density-independent) of the population growth rates.

Since culling enhances persistence when the population sizes are high, we expected the other two methods that also involve culling steps (i.e. BLC and TOC) to a degree, to be effective in terms of reducing extinction probabilities. Since the resolution in Fig 2-3 was too low for investigating this prediction, we explicitly studied the fraction of times the population sizes crashed low enough to cause concern in terms of extinction. Using all the replicate time series for each control method, we quantified the fraction of times extinction happened due to lattice effect (i.e. LEE) and demographic stochasticity (i.e. EDS). Even though the control magnitudes were set such that all methods caused an equal reduction in FI, the corresponding reductions in LEE and EDS were different, owing to how these methods changed the distribution of population sizes. ULC never allowed the population sizes to reach either the EDS or the LEE zone (Fig 5), which explains the complete lack of extinction in that control method (Fig 4A). Although, the upper ranges ($\geq 75^{th}$ percentiles) of BLC's post perturbation values were higher than those of ULC (Fig 2), they were not high enough to enter the extinction thresholds in next generation (Fig 5) and therefore caused no extinctions (Fig 4A). The same stands true for TOC in general, except for a very low frequency of extinction in the HrLk regime (Fig 5). On the other hand, the three restocking control methods that did not have a component of directly reducing the peaks in population sizes (CP, LLC and ALC) had significant number of extinctions, particularly in the two Hr regimes (Fig 4A). ALC and CP turned out to be the worst performers in terms of persistence in the HrHk and the HrLk regimes respectively. This observation was explained when we compared the EDS and LEE values in these two regimes (Fig 5A-5B). The controls in both regimes suffered primarily from LEE which both ALC and CP were able to partially ameliorate. However, CP was more effective in reducing LEE and suffered a higher fraction of EDS whereas the converse was true for ALC (Fig 5A-5B). These observations were supported by the fact that ALC consistently had higher values for the $90^{th}$ and $95^{th}$ percentiles for the post-perturbation population sizes (Fig 2A, 2B) which means they are expected to hit lower values in the next generation more often. Interestingly, in the LrLk regime, CP actually induced LEEs when there were none in the unperturbed (Fig 5C) and ultimately increased the extinction probability. LLC was found to be the most effective in terms of reducing the extinction probability in all the regimes (Fig 4A) and primarily suffered from EDS (Fig 5A-C). The overall message from all these observations is that control methods interact with growth rate and carrying capacity to alter the distribution of population sizes, which in turn determines the relative frequencies of LEE and EDS, ultimately affecting the persistence stability of populations. Since in populations with high growth rates, extinctions can only happen following a crash from a peak, methods that control the upper ranges of population sizes are better in terms of controlling extinctions for a given level of fluctuation in population sizes. However, such methods have their own share of problems in terms of applicability.

### *50% reduction in FI: Effort Magnitude*

All control methods investigated in this study involve restocking or culling of individuals. In practice, addition or removal would always incur some economic cost which can become a major factor in the choice of control methods. There are no simple ways of knowing how much the implementation of a method would cost as the figures would clearly be contingent on factors like the species under consideration, the technological know-how available etc. Therefore, following past studies [7,9], we considered the number of organisms added (or removed) to be a proxy of the economic cost involved. We assume here that all else being equal, the economic cost (and hence the undesirability of a method) is directly proportional to the number of organisms to be added or removed.

In the two Hr regimes, the three methods that involved some amount of culling (ULC, BLC, TOC) required greater effort compared to the three that relied solely on restocking (CP, LLC, ALC) (Fig 4B). This is a direct consequence of the positive-skew in the distribution of the unperturbed populations under high intrinsic growth rate (Fig 3A-3B). Ricker model of population growth ensures that whenever growth rate is high ($r > 2.0$), crashes bring the population size below the carrying capacity whereas increases in population sizes take them above $k$. Thus, in the absence of perturbation, for Hr, 50% of the points of the return map are squeezed between 0 and $k$ while the remaining 50% are spread over the interval $k$ to approximately $5k$ (Fig 3A-B). Due to this long tail of the first return map of the Ricker model [29], even small magnitudes of restocking prevent the population from reaching relatively large intervals in terms of peak sizes in the next generation. On the other hand, a much wider interval of high population sizes need to be controlled, to restrict the magnitude of the crashes in the next generation. The net result of this is that restocking methods require lesser effort to reduce the fluctuation in population sizes, compared to methods that involve culling. This line of reasoning does not fully explain the working of control methods like BLC and TOC that involve both culling and restocking. However when we explicitly looked at the magnitudes of culling and restocking for these two methods it was found that under high *r*, more effort is expended in culling than in restocking (Fig S7A-B).

The above account is expected to hold only when the population size distribution is positively skewed. This explains why under low *r* (i.e. for LrLk and LrHk), where the distribution of population sizes is symmetric, (see unperturbed in Fig 3 C-D), not much difference was noted between the effort expended in the six methods (Fig 4B). This also highlights that the effort magnitude of a method depends primarily on the growth rate of the species. It should be noted here that, by definition, the effort values of a series are scaled by the corresponding average population size (Section Methods: *Effort magnitude and Effective Population Size*). Thus, the actual values of the number of organisms to be added or removed will clearly depend on the carrying capacity. Moreover, from fig S7C and S7D, we note that for low *r*, there is more restocking for BLC whereas culling remains the dominant factor for TOC. This distinction can be important because these two processes have opposite effects on another crucial determinant of the extinction probability of a population, namely its genetic diversity.

*50% reduction in FI: Effective Population Size ($N_e$)*

A population which has lower genetic diversity can suffer from inbreeding-like deleterious effects and therefore have a greater risk of extinction [47]. In any fluctuating population, the loss of genetic diversity is accelerated each time the population size hits low numbers but is relatively unaffected by high numbers. Due to this asymmetry, average population size, which puts equal weightage on population troughs and peaks, is a poor indicator of the loss of genetic diversity.

On the other hand, effective population size (N$_e$), as measured by the harmonic mean, gets more adversely affected by small population sizes, and hence is a better index for the rate of loss of genetic diversity. In the present study, since all six methods increased the breeding size (i.e. post perturbation size, Fig 2) of the populations, the corresponding N$_e$ values were enhanced under all four regimes (Fig 4C). Since ULC led to the lowest values for both the upper (95$^{th}$ percentile) and the lower (5$^{th}$ percentile) of population size distributions (Fig 2) it was relatively less effective in terms of N$_e$, particularly under HrHk (Fig 4C). For all the other methods, both the upper and the lower ranges played a role in determining the N$_e$ and the knowledge of any one was not sufficient for prediction. In general, CP tended to lead to the highest N$_e$ under all four regimes.

*50% reduction in FI: Composite performance score*

From our results it is clear that at the level of constancy stability investigated, no single method is unambiguously superior to the others. As is often the case in biology, the "best" method was context-dependent. However, such answers mean little for most practical purposes. Therefore we computed the composite performance score with equal weightage to extinction probability, effort magnitude and effective population size (Fig 6A). Note that for this score, lower values indicate better performance and vice versa. In general, we found LLC to be the most optimal method in all four regimes (Fig 6A). ALC and BLC performed well under low *r*, whereas BLC and TOC were good under high *r* conditions.

*50% reduction in extinction probability*

The above discussion pertained to the correlates of attaining a 50% reduction in fluctuation index. However, for most practitioners of conservation, reducing the extinction probability of a population is perhaps a more pressing goal. Therefore, we repeated the entire analysis above in terms of 50% reduction in extinction probability. The exploratory analysis can be found in Fig S1-S6 whereas the persistence analogues of Figs 2-4 are Figs S8-S10. It should be noted here that the two Lr regimes (LrHk and LrLk) suffered almost no extinctions and therefore were excluded from this part of the study.

We began by looking at the effects of reducing extinction probability by 50% on the corresponding FI. The three methods that did not have a culling step (i.e. CP, LLC and ALC) had lower FI compared to the three that included a culling step (i.e ULC, BLC and TOC) (Fig S10A). Interestingly, the former set of methods was actually worse off in reducing extinction probability for a given reduction in FI (*cf* Fig 4A and Fig S10A). This once again highlights the rather complex relationship between consistency and persistence, a fact that can be observed more directly by examining Fig S1-S6.

The effort magnitude profiles were also very different particularly in the HrHk regime (*cf* Fig 4B and S10B). It can be immediately seen that it requires relatively lesser effort to reduce extinction by 50% than to achieve a similar reduction in FI. This is because the pre-images to prevent LEE or EDS are well above $k$ and hence require relatively less culling effort, whereas a population needs to be more heavily controlled to obtain a similar magnitude of reduction in FI (Table S1). An interesting manifestation of this effect can be observed in the unnaturally low effort magnitude and $N_e$ of BLC in the HrLk regime (Fig S10C). To explain this observation we recall that for BLC, multiple combinations of the upper and the lower threshold led to a 50% reduction in extinction probability and therefore arbitrarily the combination that had the lowest effort magnitude was chosen (Table S1). In the lower-threshold-upper threshold parameter space, there is a small zone of relatively high upper thresholds and really small lower thresholds that showed a 50% reduction in extinction probability. Our arbitrarily set criteria picked a point in this zone of the parameter space. Since the extinctions were mitigated by preventing crashes in the pre-perturbation population sizes, the very low value of the lower threshold did not matter in that context. However, since the $N_e$ is computed on the post-perturbation population sizes, the low values of the lower threshold led to a very small increase in post-perturbation population size, which reflected in the very small $N_e$. The very small $N_e$ of ULC in the HrHk regime arises due to a similar reason, except that in this case, there are no lower thresholds to begin with. This inability to enhance the $N_e$ proved costly for ULC and BLC in terms of the composite performance score and LLC and ALC emerged as the primary methods of choice for reducing extinction probabilities (Fig 6B).

*Caveats*

Several interesting observations emerged from this study. At least for the levels of stability investigated in this study, methods that have a culling step (ULC, BLC and TOC) are better at preventing extinctions although they are worse off in terms of reducing fluctuations in population sizes. The converse is true for methods that involve only restocking step (CP, LLC and ALC). The efficacy of control methods that incorporate both (like TOC and BLC) varies depending on whether restocking becomes the dominant force or culling. Of course, the efficacy of the methods also depend upon combination of growth rates and carrying capacities of the populations, which highlights that there is no alternative to gathering relevant biological information before the application of a control method. However, in the absence of detailed knowledge of the system, LLC (and to a lesser extent ALC), are the most optimal methods to employ (Fig 6).

Although these comparisons were obtained from simulations that incorporated several biologically realistic features, from an application point of view, there are several other caveats that need to be considered. In this study, we gave equal weightage to culling and restocking for computing the effort which, depending on the species, need not always be the case. Thus, for example, if artificial breeding of a particular species is more expensive than killing them in wild, then the entire effort calculation needs to be suitably modified. Similarly, culling and restocking have very different effects on the standing genetic variation of a population. In this study, we compute $N_e$ solely as a function of the population sizes. However, if the organisms that are used for restocking come from a different genetic stock, then for all control methods except ULC, the actual rate of loss of genetic variation would be less than what is indicated by the values of $N_e$ reported here.

In the computation of the composite performance score, we gave equal weightage to fluctuation index / extinction probability, effort magnitude and effective population size, which might not be applicable under all scenarios. We also omitted factors like frequencies of external interventions, which might become crucial under certain scenarios [26]. Any changes in these relative weightages or inclusion of more parameters for comparison can possibly lead to different conclusions in terms of relative performances of the methods. For example, we did not take into account the cost of census of the population, which actually varies among the methods. CP needs no censuses for implementation, which perhaps explains its popularity among studies that explicitly consider more realistic frameworks [48]. LLC requires lower census efforts in peak years (i.e. till the point the threshold number of animals has been sighted) and greater census efforts in lean years (when the population sizes are low, the whole population will have to be counted to decide whether the perturbations need to be implemented). The other four methods require complete census at all time intervals. It is perceivable that sometimes the cost of census might over-run the cost of implementation of the perturbations, which is evidently economically undesirable. Moreover, for many organisms like butterflies or fishes, a total count of the population might be impossible, and one would be forced to depend upon counts extrapolated from samples. Robustness of control methods towards such noisy implementations have been demonstrated for most of the methods referred here [6-7,9,19] but was not included in the present study since it is relatively difficult to quantify. However, its importance in terms of usability of a control method cannot be overstated. Finally, this study is explicitly in the context of spatially unstructured, single-species populations whereas most natural populations are expected to exist as metapopulations in multi-species assemblages which might necessitate other kinds of control strategies (e.g. [49]). Thus, a multitude of factors still need to be considered before adopting the various control methods investigated here, under field conditions. However, given that the consequences of controls going wrong can sometimes be catastrophic for ecosystems (e.g. [50]), the kind of comparison that we attempt here will be a crucial part of translating theory into practice.

# Acknowledgements

We thank Amitabh Joshi and Pratha Sah for helpful comments on the manuscript.

**Figure Legends**

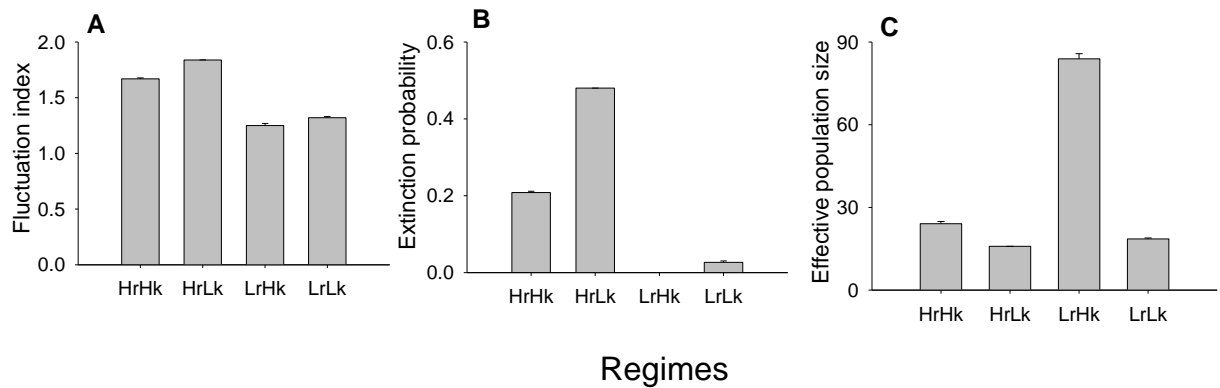

**Figure 1. Dynamics of unperturbed populations in four regimes.** (A) Average fluctuation index, (B) average extinction probability, and (C) average effective population size of the unperturbed populations calculated over 50 replicate runs for each of the four regimes. In this and all subsequent figures, HrHk stands for High *r* (= 4.0) and High *k* (= 300), LrLk denotes a combination of Low *r* (= 2.8) and Low *k* (= 60), and so on for the other two regimes. Error bars represent SEM. The two Hr regimes had higher FI and extinction probability compared to the two Lk regimes. Although the effective population sizes of the two Hr regimes were comparable to that of the LrLk regime, this was an artifact of the simulation protocol (see Section Results and Discussion: Dynamics of unperturbed populations for more details).

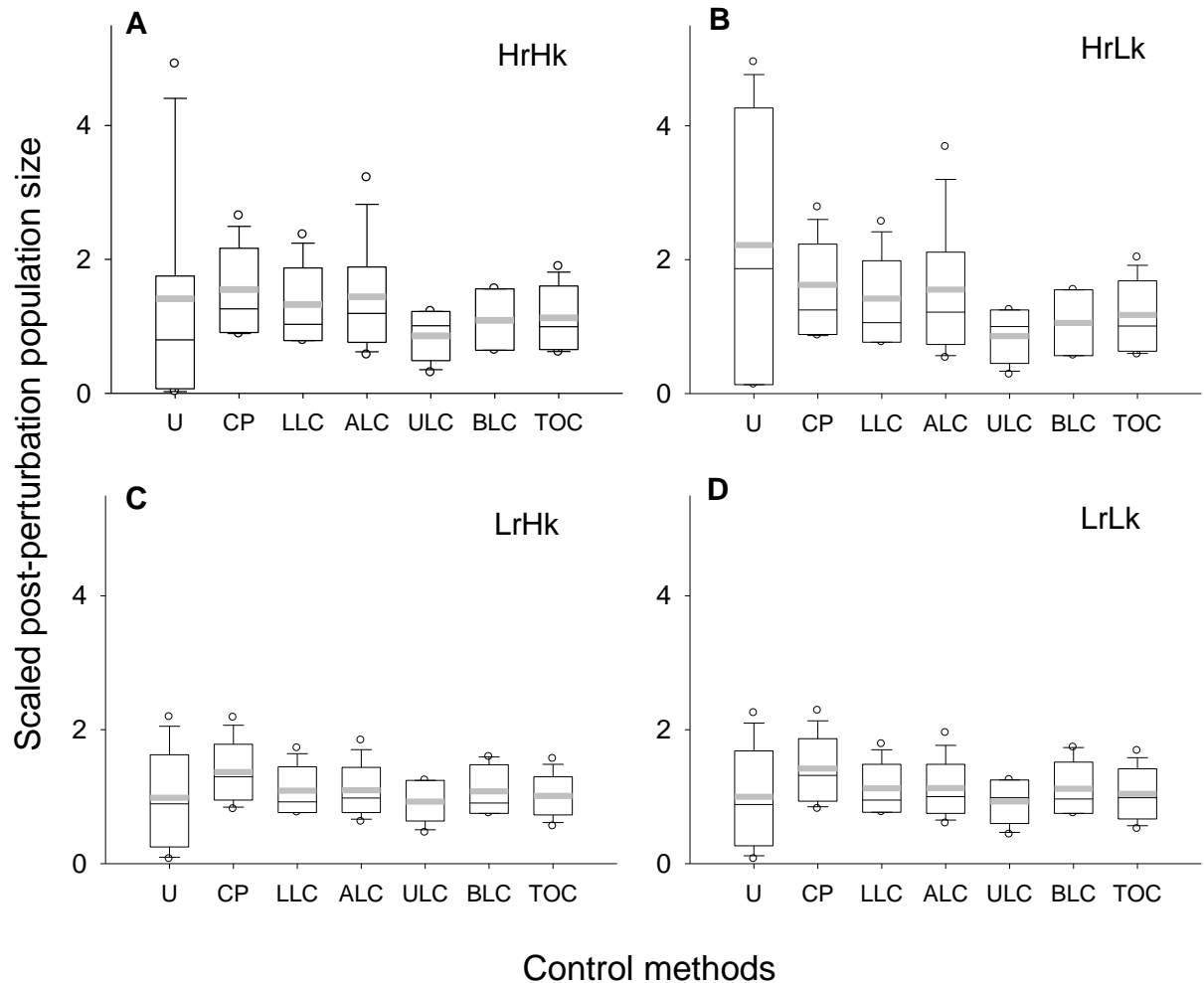

**Figure 2**. **Box-plots of the post-perturbation population size distributions for 50% reduction in Fluctuation Index**. A, B, C, and D correspond to the HrHk, HrLk, LrHk and LrLk regimes respectively. Thick grey lines = means, thin black lines in the box = medians. Lower and upper limits of the box represent 25$^{th}$ and 75$^{th}$ percentiles, lower and upper whiskers denote 10$^{th}$ and 90$^{th}$ percentiles while the lower and upper dots stand for 5$^{th}$ and 95$^{th}$ percentiles. U stands for the unperturbed population while the other abbreviations denote the six methods. All the values have been scaled by the corresponding carrying capacity (*k*) of the regime to facilitate comparison. ULC always had the lowest post-perturbation population size, since it did not involve any restocking.

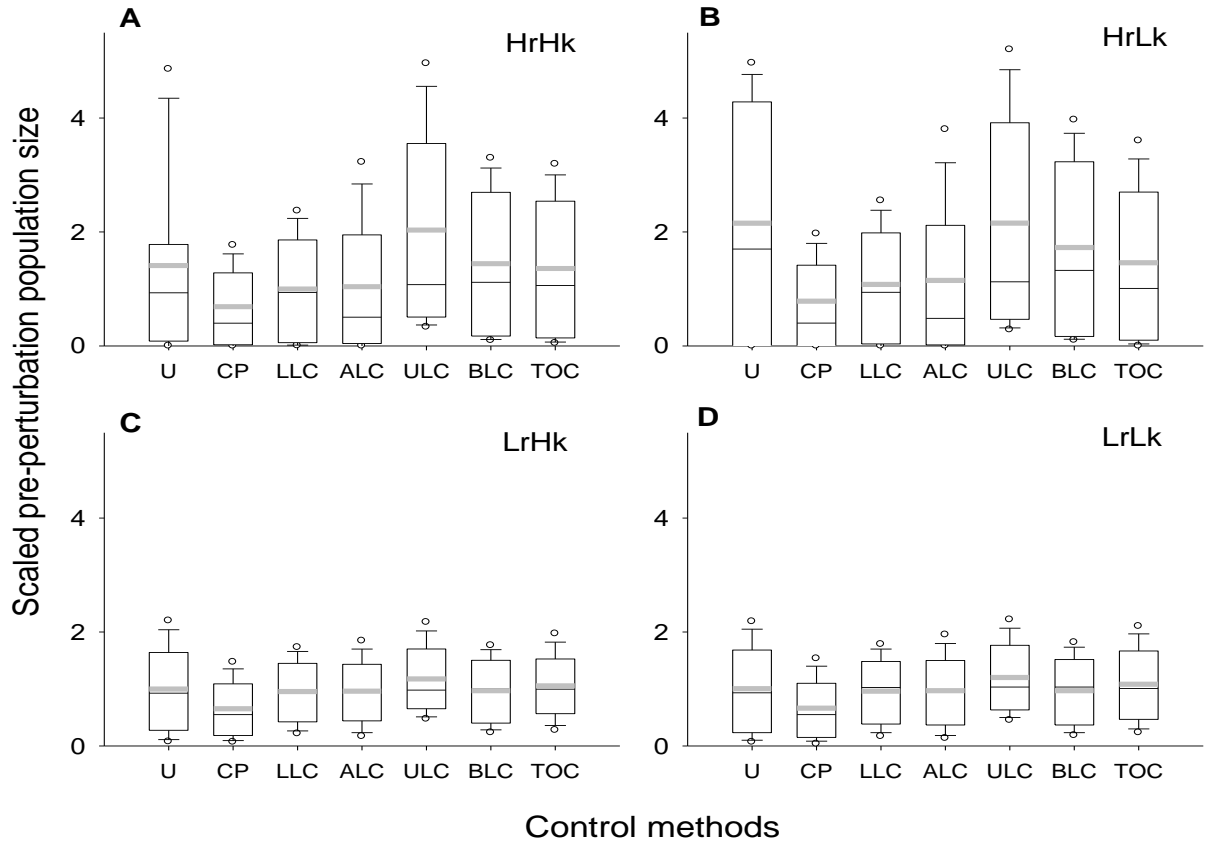

**Figure 3. Box-plots of the pre-perturbation population size distributions for 50% reduction in Fluctuation Index**. A, B, C, and D correspond to the HrHk, HrLk, LrHk and LrLk regimes respectively. Thick grey lines = means, thin black lines in the box = medians. Lower and upper limits of the box represent 25$^{th}$ and 75$^{th}$ percentiles, lower and upper whiskers denote 10$^{th}$ and 90$^{th}$ percentiles while the lower and upper dots stand for 5$^{th}$ and 95$^{th}$ percentiles. U stands for the unperturbed population while the other abbreviations denote the six methods. All the values have been scaled by the corresponding carrying capacity (*k*) of the regime to facilitate comparison. ULC had the highest pre-perturbation population size, which indicated that it would also have the least amount of extinctions. Note that the size distribution of the unperturbed (U) populations is asymmetric in the two Hr regimes (A,B), but not in the two Lr regimes (C,D). This has consequences for the corresponding effort magnitudes.

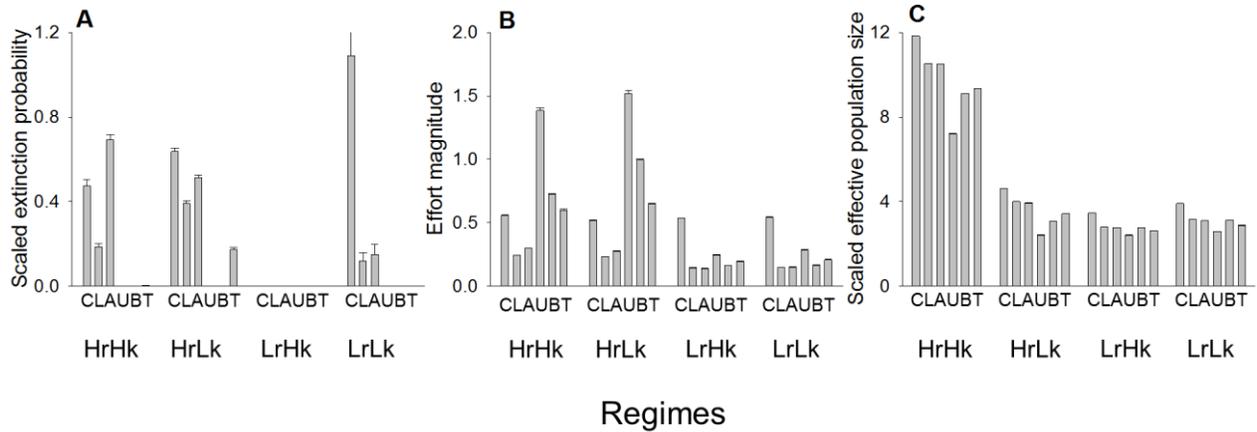

**Figure 4. Comparison of the six methods for 50% reduction in Fluctuation Index.** C = CP, L = LLC, A = ALC, U = ULC, B = BLC, T = TOC. (A) Average extinction probability. Note that there were no extinctions in the LrHk regime. The absence of a method within a regime indicates no extinctions. (B) Average effort magnitude corresponding to the six methods in four regimes. (C) Average effective population size corresponding to the six methods in four regimes. In panels A and C, each value has been scaled by the average value of the unperturbed population in that regime. In general, the methods that involve a culling step, i.e. ULC, BLC and TOC, are better at reducing extinction probability. However, they have higher effort magnitudes in the two Hr regimes. All methods were effective in increasing the effective population size compared to the controls.

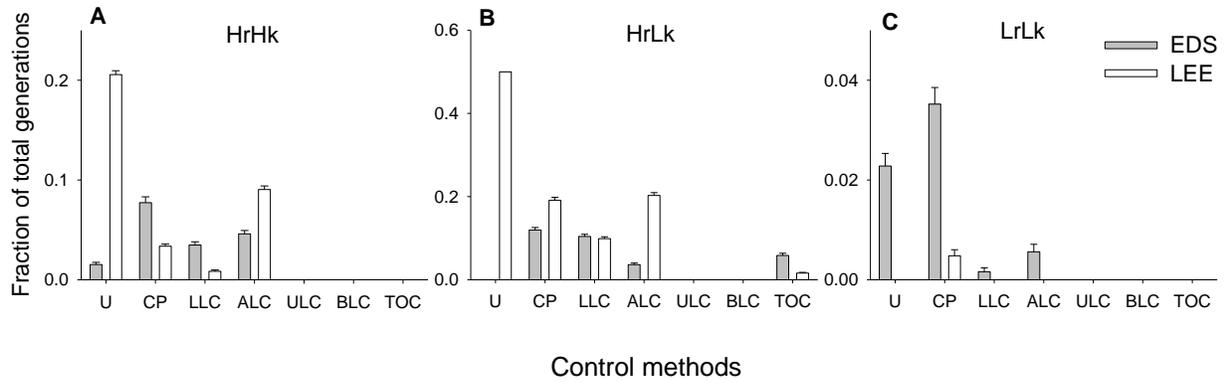

**Figure 5. Lattice Effect Extinctions Vs Demographic Stochasticity Extinctions for 50% reduction in fluctuation index.** Average fraction of times extinction occurred due to population size falling below 0.5 (LEE) or between 1 and 3 (EDS). (A) HrHk, (B) HrLk and (C) LrLk. No extinction occurs in LrHk regime and hence it has been omitted here. The extinctions were scored over 50 generations for 50 replicate runs for each method in each regime. Note the differences in the Y-axis scales in the three figures. For a similar reduction in fluctuation index, the relative frequencies of LEE and EDS varied between the control methods.

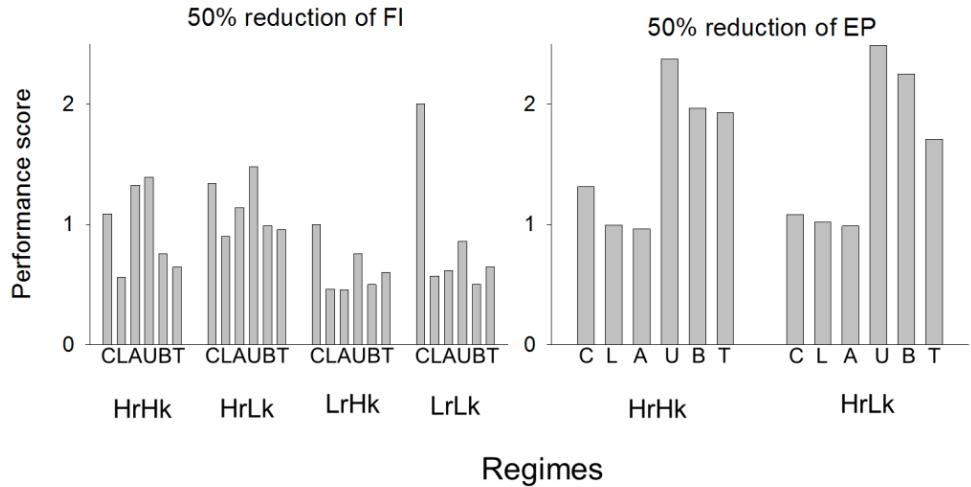

**Figure 6. Composite performance score for comparison between the control methods.** C = CP, L = LLC, A = ALC, U = ULC, B = BLC, T = TOC. A) For 50% reduction in FI. (B) For 50% reduction in extinction probability. Lower values indicate better performance. Although no method was clearly superior in all the regimes, in general, LLC performed the best under most circumstances and ALC was the second best.

**Table 1. Details of the six control methods compared in this study***

| Sl. No. | Control Method | Mathematical expression | Control parameter constants | Control parameter range(s) for Fig S1-S6 | Step size |
|---|---|---|---|---|---|
| 1. | Constant Pinning (CP) | $a_t = b_t + p$ | Pin (p) | 1 to k-1 | 1 |
| 2. | Lower Limiter Control (LLC) | $a_t = max\ [b_t, h]$ | Lower limit (h) | 1 to k-1 | 1 |
| 3. | Adaptive Limiter Control (ALC) | $a_t = max\ [b_t, c \times a_{t-1}]$ | ALC intensity (c) | 0.05 to 0.95 | 0.05 |
| 4. | Upper Limiter Control (ULC) | $a_t = min\ [b_t, H]$ | Upper limit (H) | k+1 to 3k | 1 |
| 5. | Both Limiter Control (BLC) | $a_t = max\ [h, min[\ b_t, H]]$ | Lower limit (h) | 1 to k-1 | 1 |
| | | | Upper limit (H) | k+1 to 3k | 1 |
| 6. | Target Oriented Control (TOC) | $a_t = max\ [0, c_d \times T + (1 - c_d) \times b_t)]$ | Target, T | k | NA |
| | | | $c_d$ | 0.05 to 0.95 | 0.05 |

*$b_t$ and $a_t$ are the population sizes before and after perturbation in the $t^{th}$ generation, such that $b_{t+1} = f(a_t)$, where $f$ stands for the population recruitment function (Ricker model, in this study). For TOC, we fixed the target at $k$, since prior work indicates that this is the optimal strategy for this method in terms of minimizing interventions [6]. For BLC, $H > h$. NA denotes not applicable.

Supplementary materials

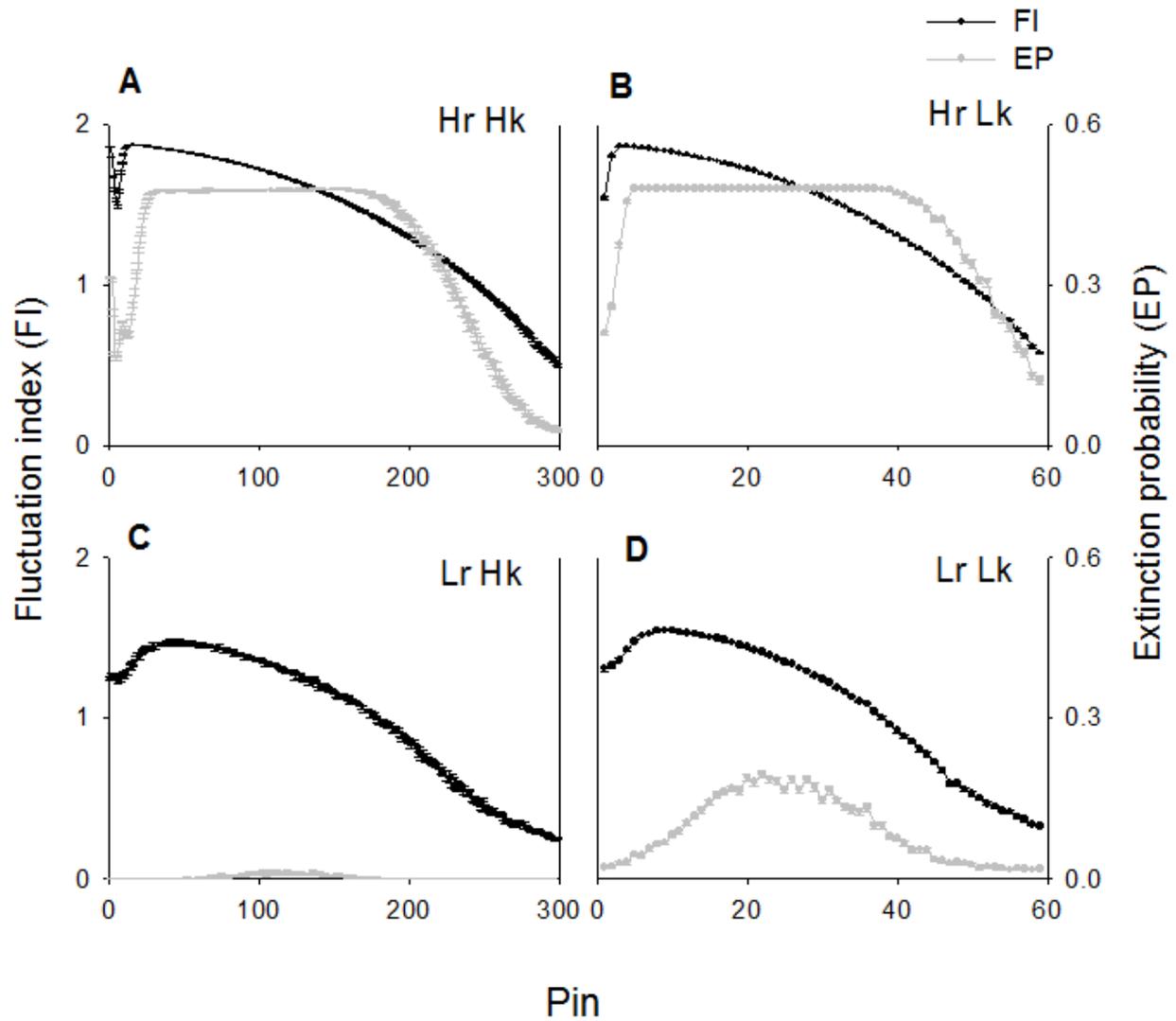

Figure S1: Average fluctuation index (± SE; black) and extinction probability (± SE; grey) for different levels of immigration under the Constant Pinning (CP) method. A, B, C and D represent HrHk, HrLk, LrHk, and LrLk regimes respectively. Error bar at each point is too small to be visible clearly.

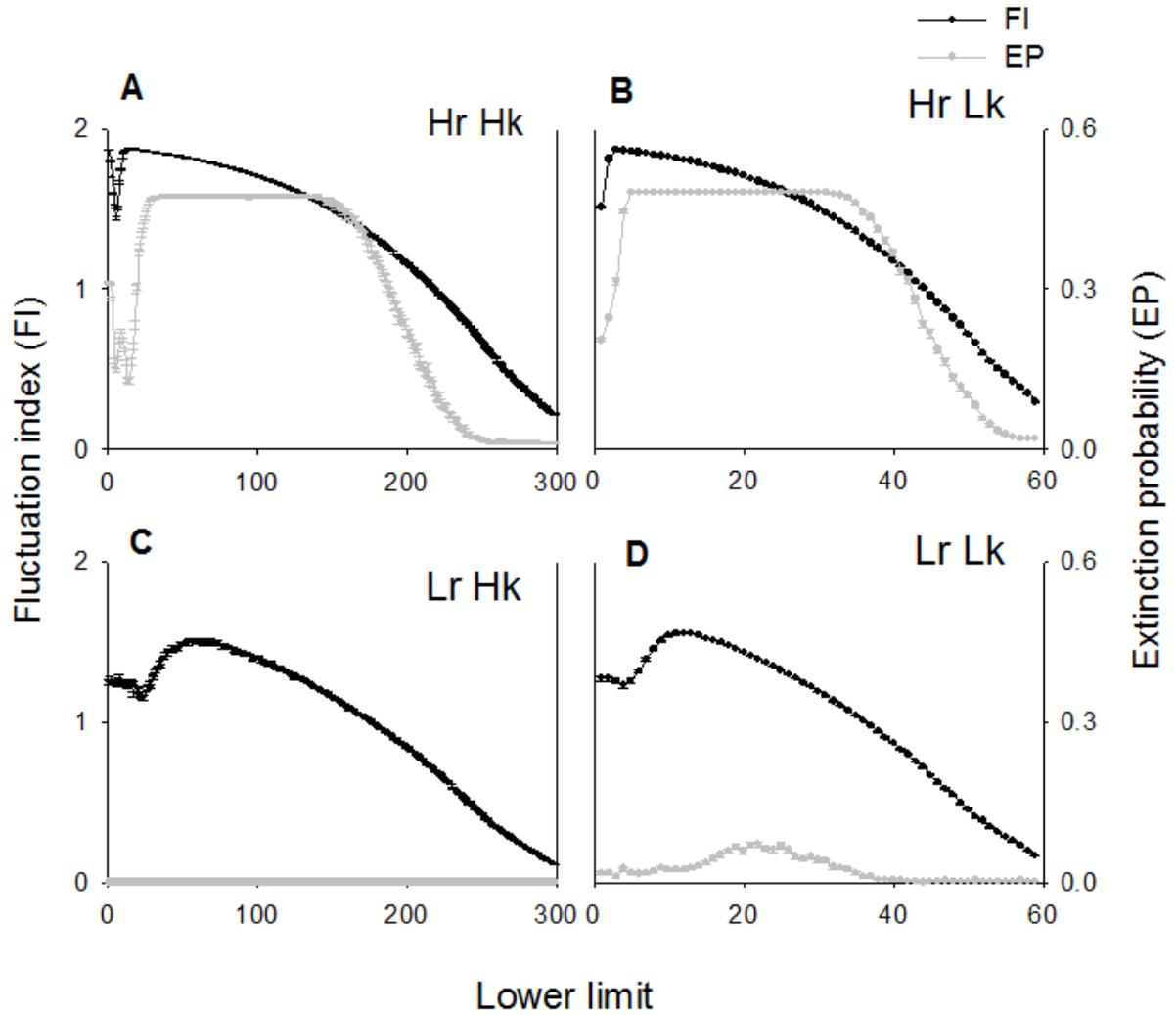

Figure S2: Average fluctuation index (± SE; black) and extinction probability (± SE; grey) for different levels of lower threshold under the Lower Limiter Control Method (LLC). A, B, C and D represent HrHk, HrLk, LrHk, and LrLk regimes respectively. Error bar at each point is too small to be visible clearly.

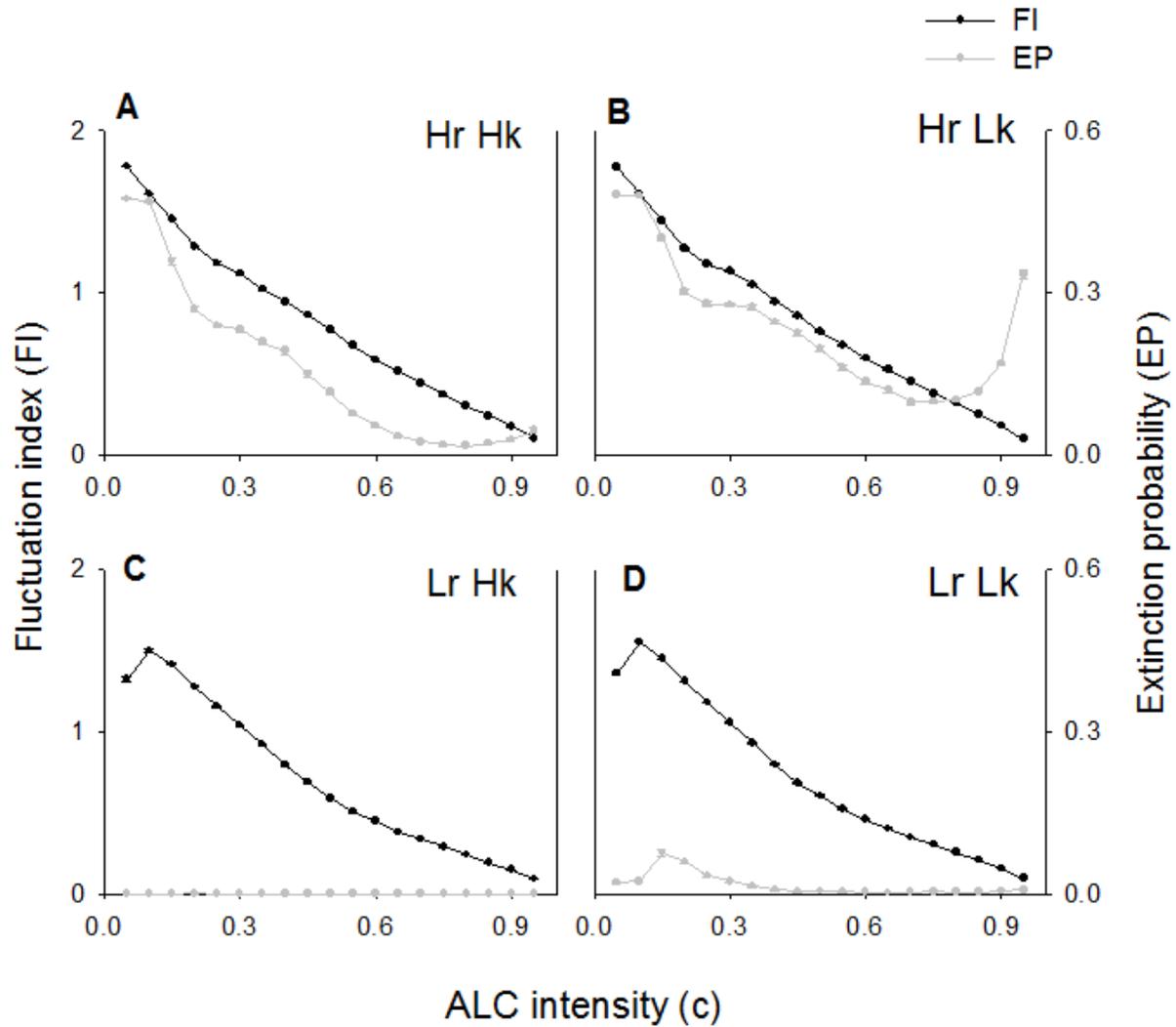

Figure S3: Average fluctuation index (± SE; black) and extinction probability (± SE; grey) for different levels of ALC intensity (*c*) under the Adaptive Limiter Control Method (ALC). A, B, C and D represent HrHk, HrLk, LrHk, and LrLk regimes respectively. Error bar at each point is too small to be visible clearly.

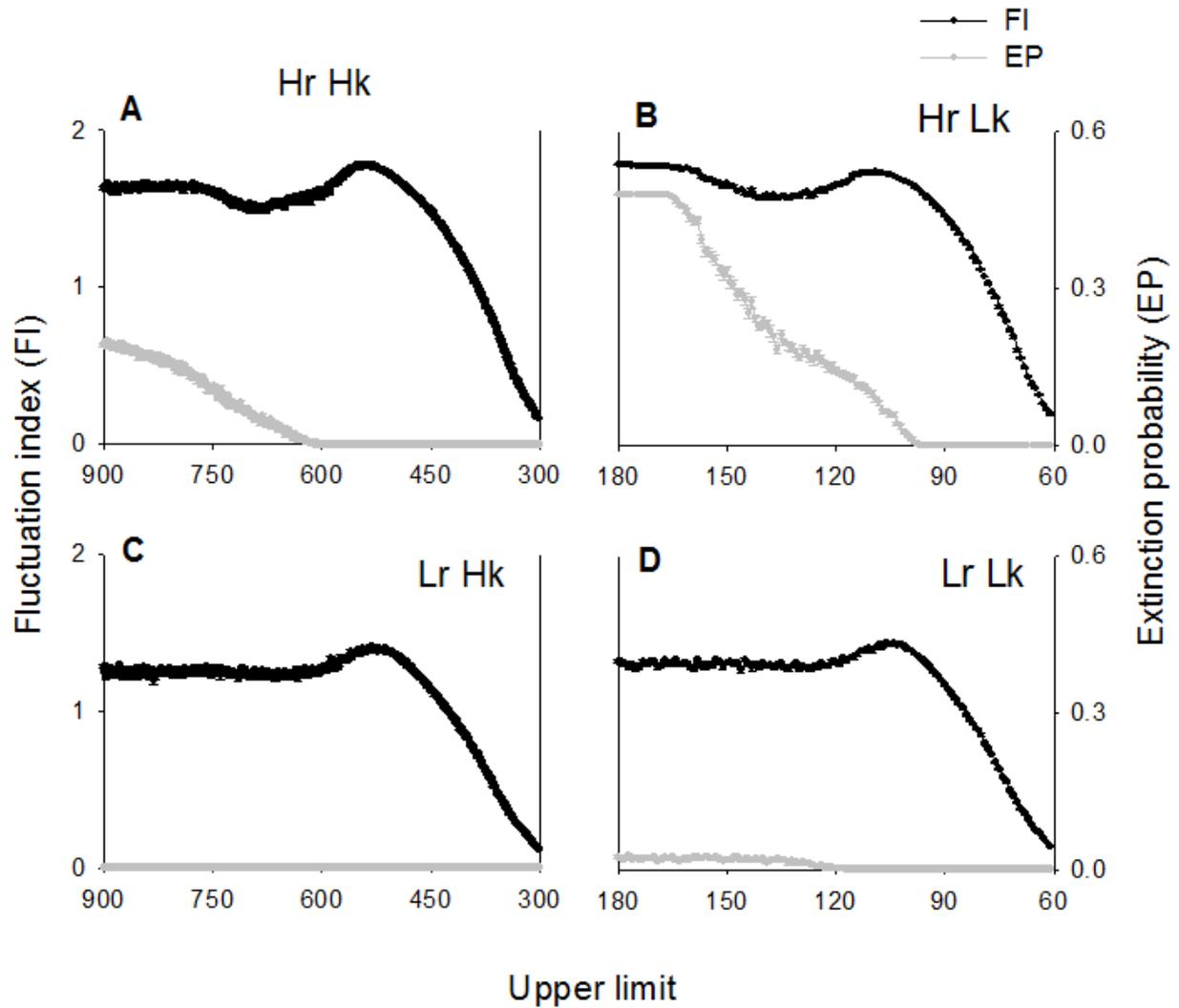

Figure S4: Average fluctuation index (± SE; black) and extinction probability (± SE; grey) for different levels of upper threshold under the Upper Limiter Control Method (ULC). A, B, C and D represent HrHk, HrLk, LrHk, and LrLk regimes respectively. Error bar at each point is too small to be visible clearly.

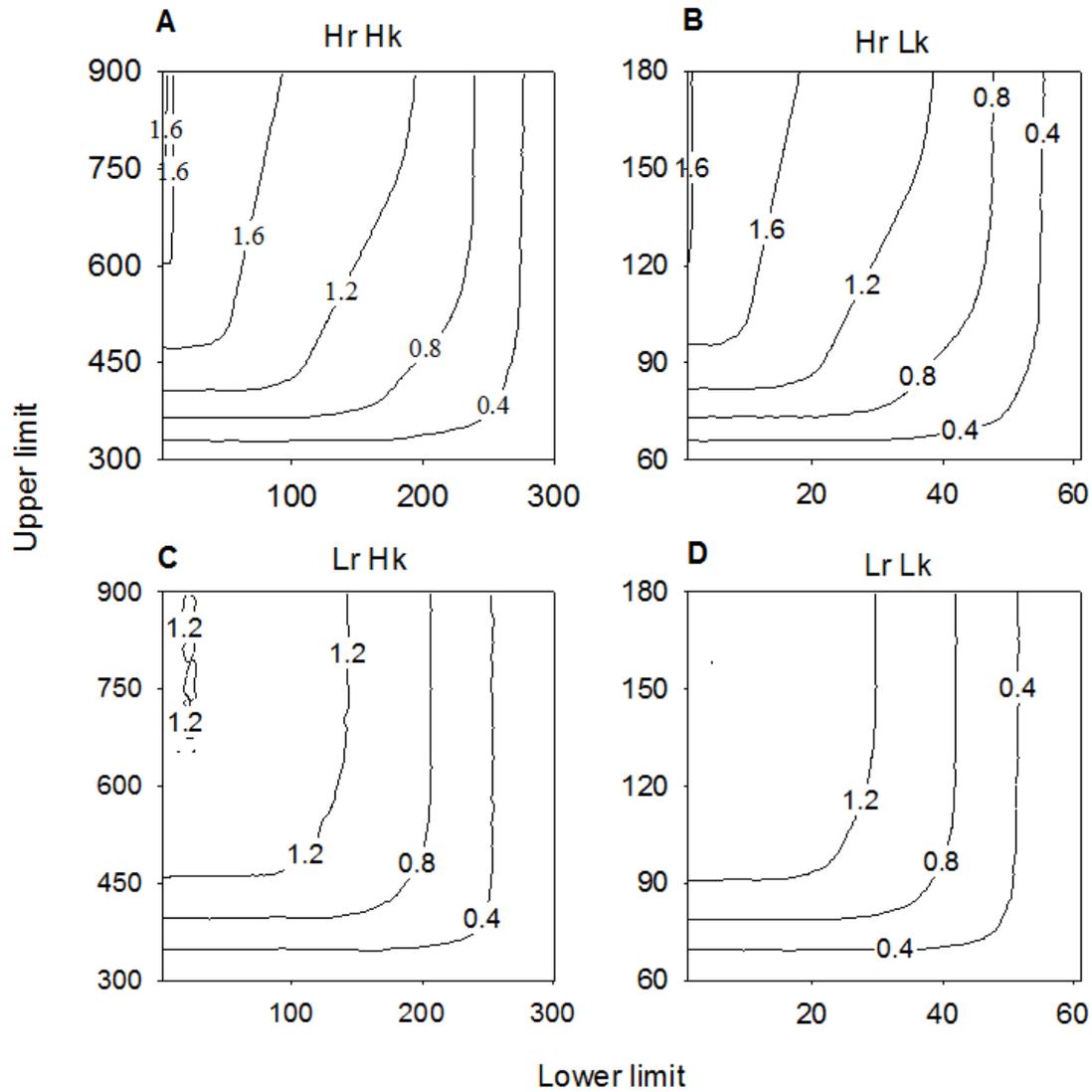

**Figure S5.1:** Average fluctuation index for different combinations of upper and lower threshold under the Both Limiter Control Method (BLC). A, B, C and D represent HrHk, HrLk, LrHk, and LrLk regimes respectively.

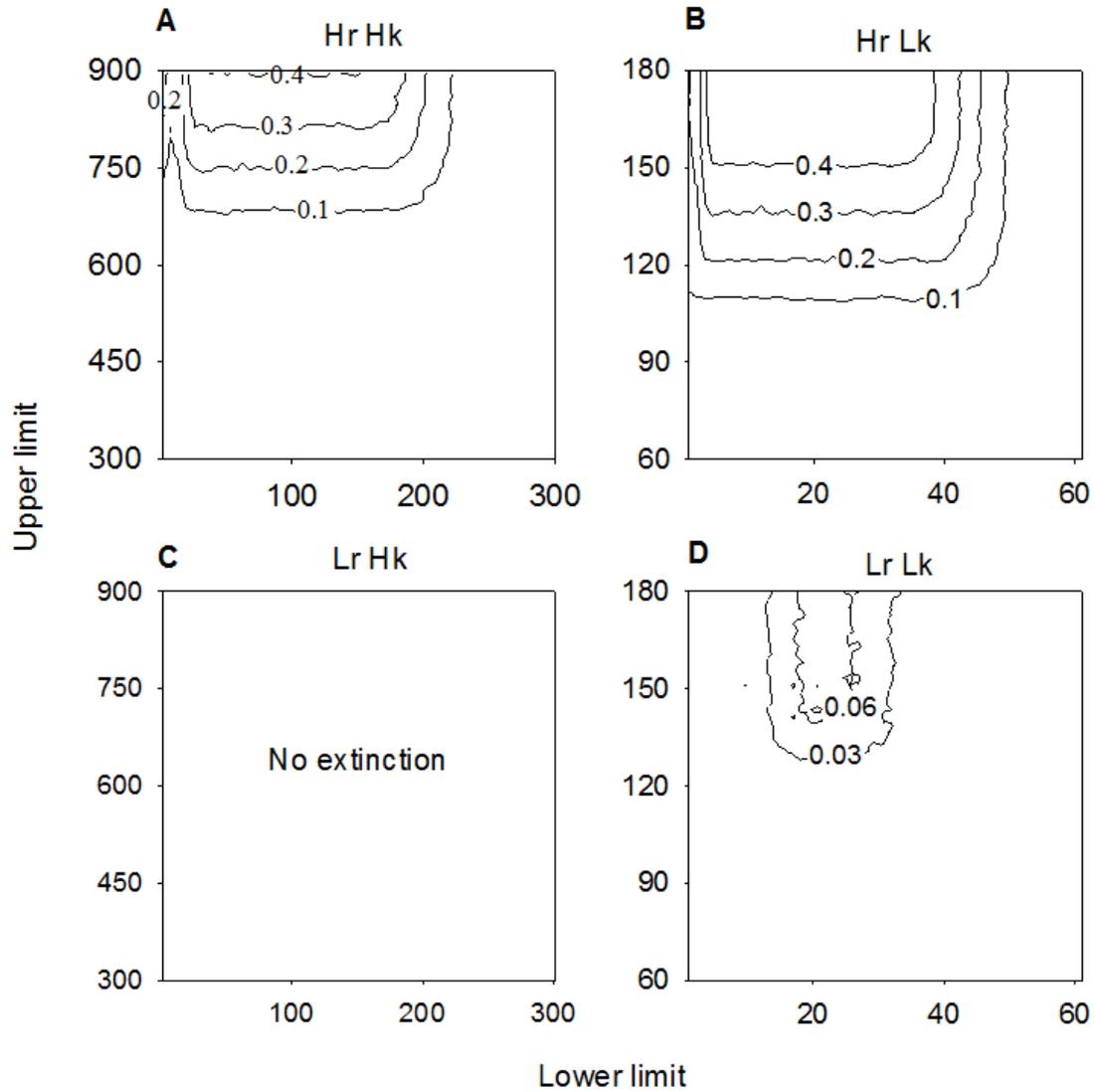

**Figure S5.2:** Average extinction probability for different combinations of upper and lower threshold under the Both Limiter Control Method (BLC). A, B, C and D represent HrHk, HrLk, LrHk, and LrLk regimes respectively.

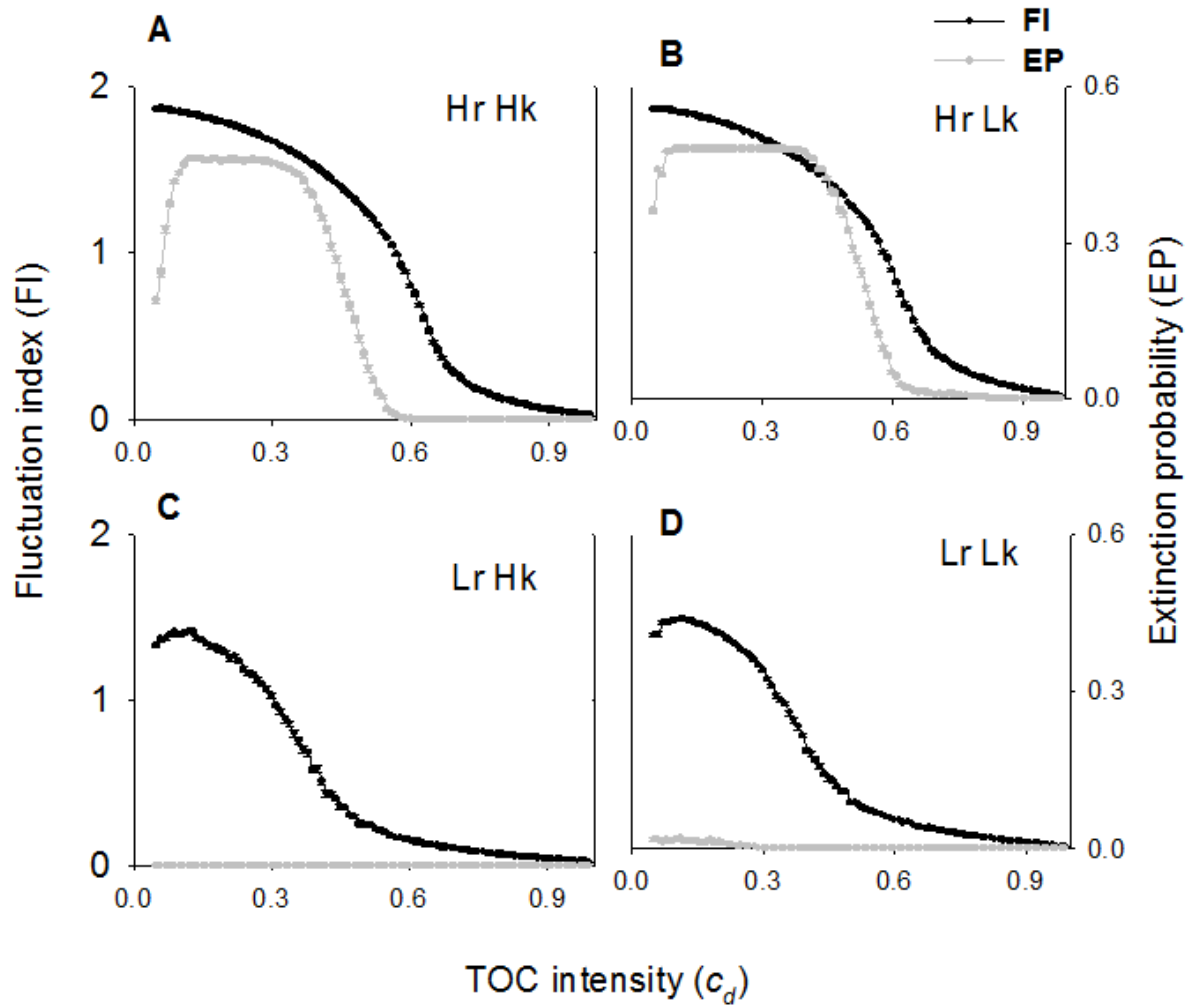

**Figure S6:** Average fluctuation index (± SE; black) and extinction probability (± SE; grey) for different levels of TOC intensity ($c_d$) under the Target Oriented Control Method (TOC). A, B, C and D represent HrHk, HrLk, LrHk, and LrLk regimes respectively. Error bar at each point is too small to be visible clearly.

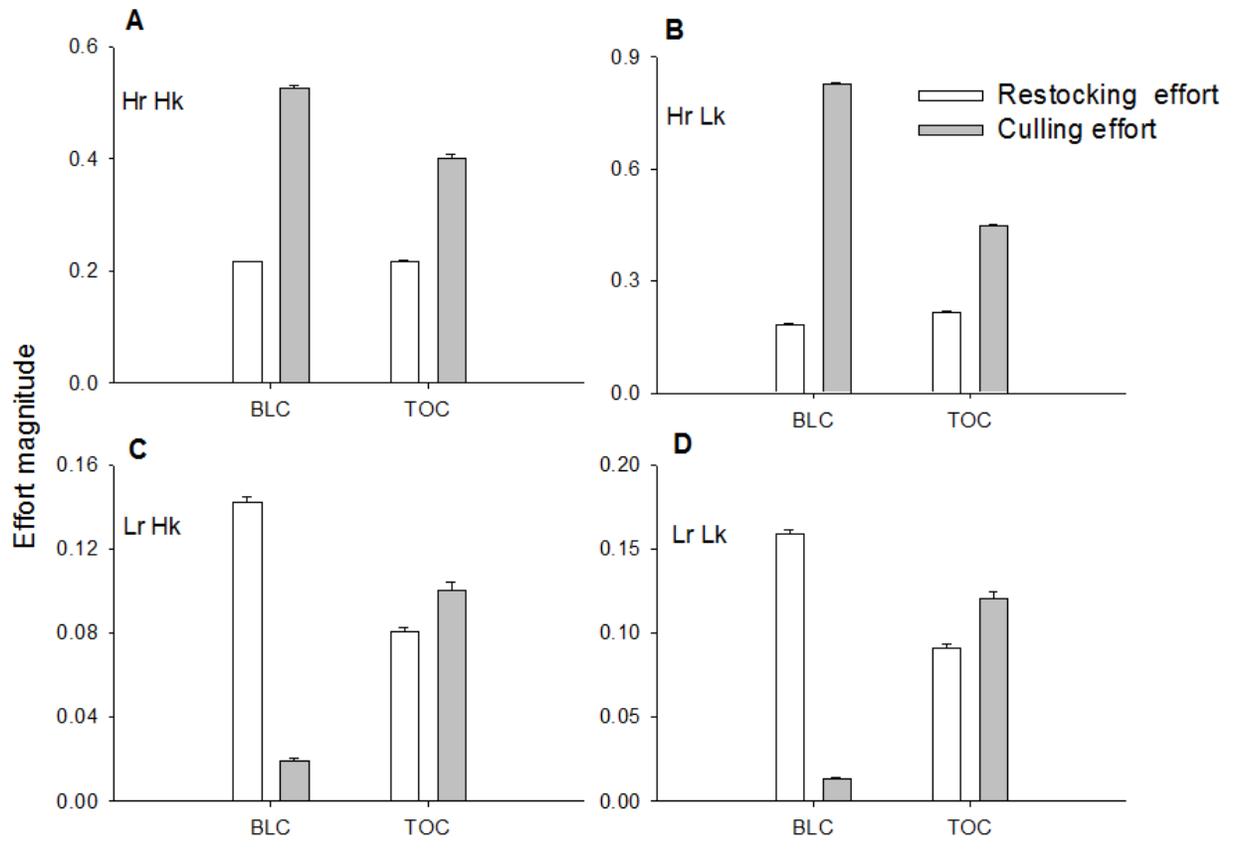

**Figure S7. Restocking and culling efforts for BLC and TOC.** Panel A, B, C and D represent the regimes HrHk, HrLk, LrHk and LrLk respectively.

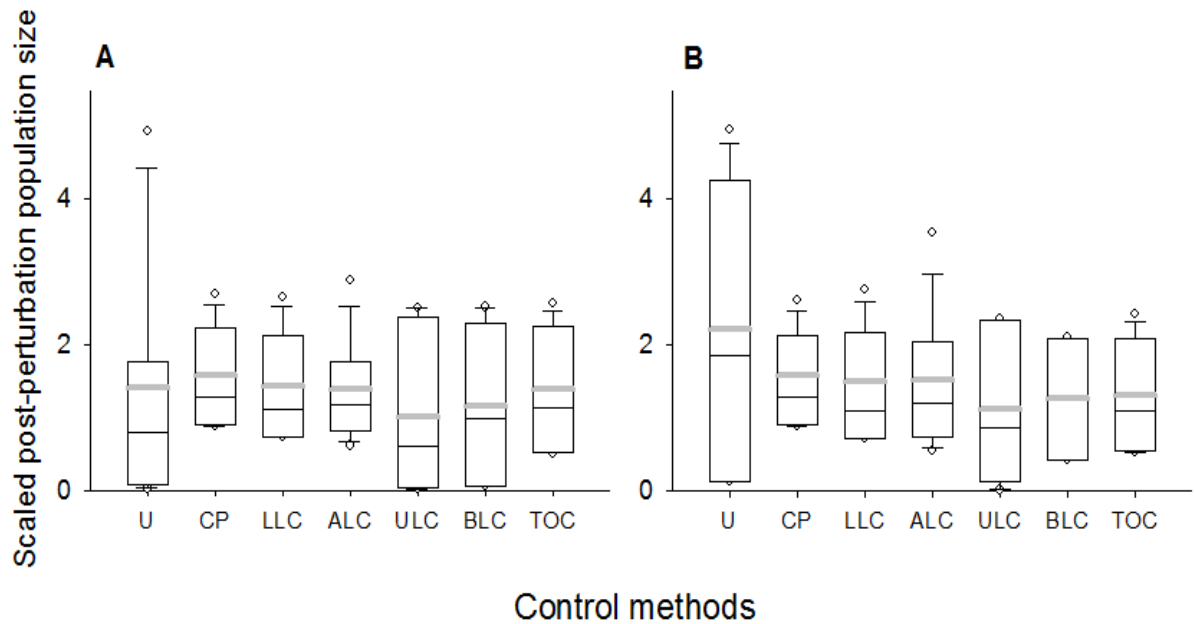

Figure S8: **Box-plots of the post-perturbation population size distributions for 50% reduction in Extinction Probability**. A, B, C, and D correspond to the HrHk, HrLk, LrHk and LrLk regimes respectively. Thick grey lines = means, thin black lines in the box = medians. Lower and upper limits of the box represent 25$^{th}$ and 75$^{th}$ percentiles, lower and upper whiskers denote 10$^{th}$ and 90$^{th}$ percentiles while the lower and upper dots stand for 5$^{th}$ and 95$^{th}$ percentiles. U stands for the unperturbed population while the other abbreviations denote the six methods. All the values have been scaled by the corresponding carrying capacity ($k$) of the regime to facilitate comparison across regimes.

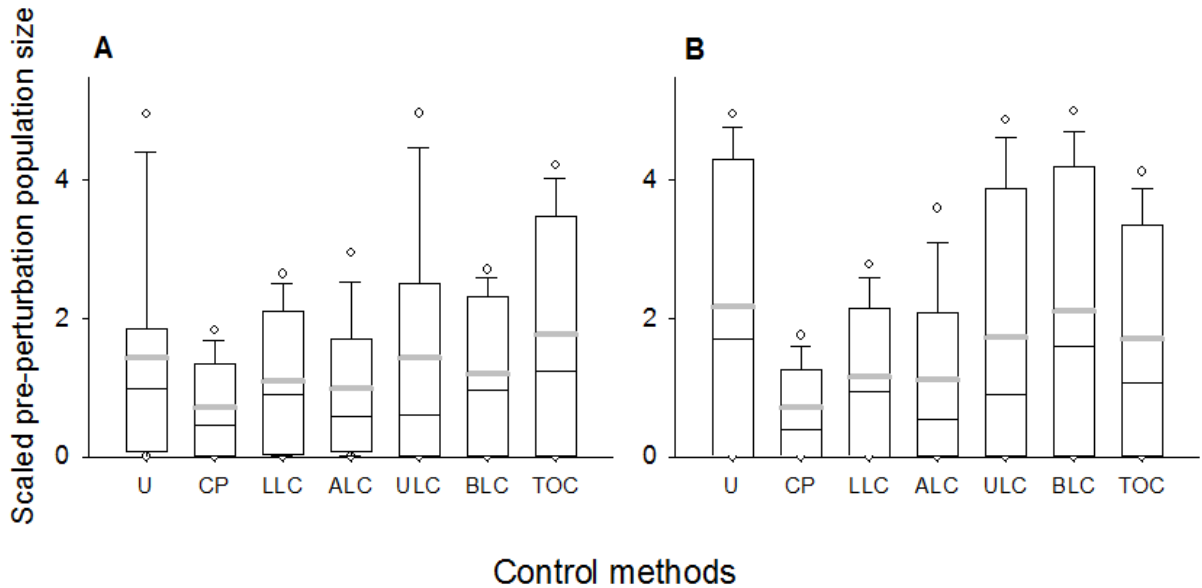

**Figure S9: Box-plots of the pre-perturbation population size distributions for 50% reduction in Extinction Probability**. A, B, C, and D correspond to the HrHk, HrLk, LrHk and LrLk regimes respectively. Thick grey lines = means, thin black lines in the box = medians. Lower and upper limits of the box represent 25$^{th}$ and 75$^{th}$ percentiles, lower and upper whiskers denote 10$^{th}$ and 90$^{th}$ percentiles while the lower and upper dots stand for 5$^{th}$ and 95$^{th}$ percentiles. U stands for the unperturbed population while the other abbreviations denote the six methods. All the values have been scaled by the corresponding carrying capacity ($k$) of the regime to facilitate comparison across regimes.

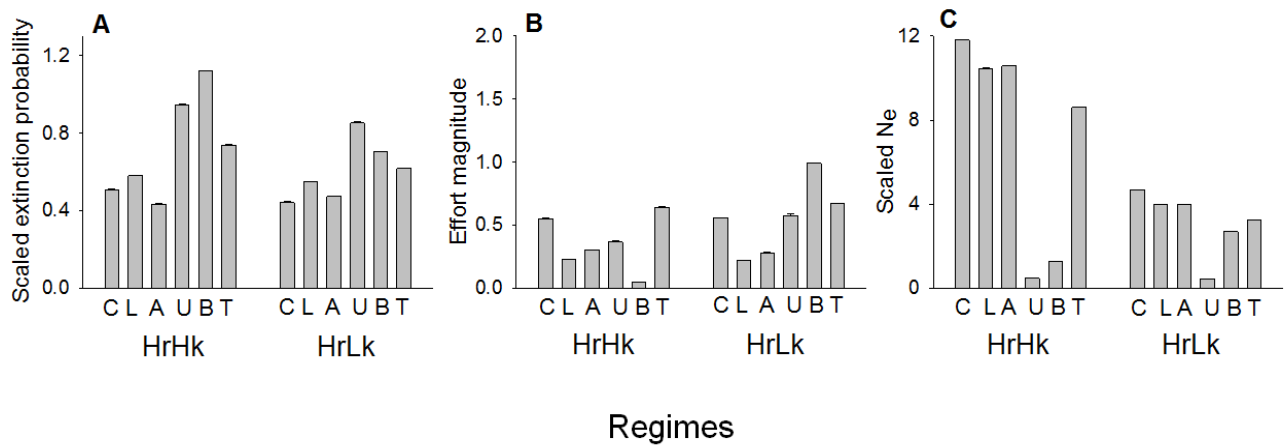

**Figure S10: Comparison of the six methods for 50% reduction of extinction probability.** Note that the unperturbed populations in LrHk and LrLk regimes suffered almost no extinctions and were excluded. (A) Average fluctuation index (± SE). (B) Average effort magnitude (± SE). (C) Average effective population size (± SE). In panels A and C, each value has been scaled by the average value of the unperturbed population in that regime.

**Table S1. Parameter values of the six control methods leading to 50% enhancement of stability***

|  | Regimes | CP | LLC | ALC | ULC | BLC | | TOC | |
|---|---|---|---|---|---|---|---|---|---|
|  |  | p | h | c | H | *h* | *H* | T | $c_d$ |
| **50% reduction in Fluctuation Index** | HrHk | 266 | 236 | 0.47 | 367 | 193 | 468 | 300 | 0.6 |
|  | HrLk | 52 | 46 | 0.42 | 75 | 34 | 93 | 60 | 0.59 |
|  | LrHk | 225 | 229 | 0.49 | 373 | 225 | 478 | 300 | 0.4 |
|  | LrLk | 47 | 46 | 0.47 | 75 | 45 | 104 | 60 | 0.4 |
| **50% reduction in Extinction Probability** | HrHk | 264 | 222 | 0.52 | 751 | 15 | 753 | 300 | 0.51 |
|  | HrLk | 54 | 44 | 0.44 | 141 | 26 | 126 | 60 | 0.54 |

*Note that these values represent the thresholds of population sizes / values of control parameters set for attaining 50% reduction in FI or extinction probability, and not the number of organisms to be added or subtracted. In case of BLC, since multiple parameter combinations satisfied the criteria of 50% reduction in FI and extinction probability, we chose the combination with the lowest effort magnitude.